\documentstyle[12pt]{article}
\input epsf
\begin{document}

\thispagestyle{empty}
\rightline{NSF-ITP-94-96}
\rightline{hep-th/9409168}

\vskip 3.0 cm

\begin{center}
{\large\bf ON THE NONPERTURBATIVE CONSISTENCY\\
OF $d=2$ STRING THEORY}
\break

\vskip 2.0 cm
{\bf Joseph Polchinski}\footnote{Electronic address:
joep@sbitp.itp.ucsb.edu}

\vskip 0.7 cm
\sl
Institute for Theoretical Physics  \break
University of California  \break
Santa Barbara, CA 93106-4030  \break

\end{center}
\vskip 1.4 cm
\rm

\begin{quote}
{\bf ABSTRACT:} An infinite number of distinct $d=1$ matrix
models reproduce the perturbation theory of $d=2$ string theory.
Due to constraints of causality, however, we argue that
none of the existing constructions gives a
consistent nonperturbative definition of the $d=2$ string.
\end{quote}
\vskip1cm

\normalsize

\newpage

The search for a nonperturbative formulation of string theory
remains a key open problem.  String theories in two or fewer
spacetime dimensions, which are exactly solvable by matrix model
methods, have been valuable model systems; for reviews, see
ref.~\cite{mmreviews}.  One notable feature of these models is the
large-order growth of perturbation theory, more rapid than in field
theory and so corresponding to larger nonperturbative effects.  This
growth was subsequently found to be generic to string theory in higher
dimensions\cite{Shenker}.

Another notable,
and puzzling, feature is a very large non-perturbative ambiguity.
There is an infinite-parameter set of distinct $d=1$ matrix models,
all of which reproduce the perturbation theory for the $d=2$ string
and all of which are consistent quantum theories\cite{MPR}.  This has
been interpreted as the existence of an infinite number of
non-perturbative parameters analogous to the $\theta$-parameter of
QCD.

In fact, we will argue that the situation is quite the
opposite---that none of the proposed non-perturbative definitions
gives rise to a consistent string theory, and that the problem of
finding {\it any} consistent non-perturbative formulation of the
$d=2$ string remains open.  This result is a simple but unexpected
consequence of a recent study of spacetime gravity in these
models\cite{NP}.

To review: after diagonalizing and
double-scaling, the matrix model becomes a theory of free fermions
in an inverted harmonic oscillator potential\cite{BIPZ,d1refs}.  In
terms of a second-quantized fermion $\psi(x,t)$ with $x$ the rescaled
eigenvalue, the Hamiltonian is
\begin{equation}
H = \frac{1}{2} \int_{-\infty}^\infty dx \,\Bigl\{ \partial_x
\psi^\dagger \partial_x \psi - x^2 \psi^\dagger \psi \Bigr\}\ .
\end{equation}
The string tachyon, which is the only propagating degree of
freedom in $d=2$, corresponds to the collective motion of the
Fermi surface\cite{fferm}.  The static solutions are given by filling
the Fermi sea on one side of the barrier, say the left, to an energy
$-\overline\mu$, below the maximum of the potential.  In string
perturbation theory, at fixed energy and fixed number of external
particles, the other side of the barrier is irrelevant.
The asymptotic region~$\phi \to -\infty$ of the
string Liouville coordinate corresponds to the asymptotic region
$x \to -\infty$ of eigenvalue space.

Fermions will tunnel through the potential barrier; this is the
anomalously large non-perturbative effect mentioned above.  To
define the theory non-perturbatively we must make a prescription
for the state on the other side of the barrier.  One class of
theories (type~I in the terminology of ref.~\cite{MPR}) eliminates
the second asymptotic region by modifying the potential.  For
example, a sharp infinite barrier, $V(x) = \infty$ for $x > A$,
leaves the perturbation theory unaffected for any $A \geq
0$.  So does any other modification such that $V(x)$ is
$-\frac{1}{2} x^2$ for
$x < 0$ and rises to infinity as $x \to \infty$.  All such
modifications give the same perturbation theory, and
all produce a manifestly unitary quantum theory within the Hilbert
space of incoming and outgoing fermions in the left asymptotic
region (or the bosonized equivalent).  Nevertheless, we will argue
that {\it no} matrix model of type~I corresponds to a consistent
string theory. An alternate approach to defining the theory is to
leave the potential unmodified and work within the larger space of
states of two asymptotic regions (type II), for example by filling
the right Fermi sea to the same level $-\overline\mu$.  We will argue
that at least the naive implementation of this idea is
inconsistent.

Since the type~I matrix models are certainly consistent quantum
theories in their own right, why do we assert that they correspond
to inconsistent string theories?  The additional consistency
condition we impose is {\it causality.}  For example, the
non-perturbative dynamics must conserve the gravitational mass,
which can be measured by scattering experiments arbitrarily far
into the asymptotic region.  That this is a nontrivial condition
follows from the rather convoluted way that gravitational physics
is encoded in the matrix model\cite{NP}.
In fact, the $d=2$ string must satisfy an infinite set of such
causality conditions.

Rather than gravitational scattering, we will focus on the simplest
process that leads to a causality condition, namely $2 \to 1$ bulk
tachyon scattering.  That is, a pair of
incoming tachyons have an amplitude to scatter into an outgoing tachyon
before reaching the Liouville wall.  This amplitude is
nonzero because the operator product of three tachyon vertex operators
contains the identity,
but does not occur in the matrix model because a small free-fermi pulse
will always travel to the turning point $x = - \sqrt{2\overline\mu}$
before reflecting.  The difference arises because of the nonlocal
relation between the string tachyon and the collective field of the
matrix model\cite{legpoles,NP}.

Let us briefly review how this comes about.  Comparison of the
matrix model and string S-matrices shows that the respective tachyon
fields are related in the asymptotic region by the so-called `leg pole' factor
\begin{equation}
{\cal S}( \phi ,t ) \ =\ \biggl( \frac{\pi}{2} \biggr)^{-
\partial_\phi / 4}
\frac{\Gamma(-\partial_\phi)}{\Gamma(\partial_\phi)}
\overline {\cal S}(\phi, t), \label{renorm}
\end{equation}
where the bar identifies the matrix model field.  The string
tachyon $\cal S$ is a function of $\phi$ and $t$, while
the matrix model tachyon $\overline{\cal S}$ is defined as a
function of
$q$ (logarithm of the eigenvalue, defined below) and
$t$.  Eq.~(\ref{renorm}) relates the $\phi$-dependence of $\cal S$
to the $q$-dependence of $\overline{\cal S}$.
This transformation is nonlocal, and gives rise to
bulk scattering as shown in figure~1.
\begin{figure}
\begin{center}
\leavevmode
\epsfbox{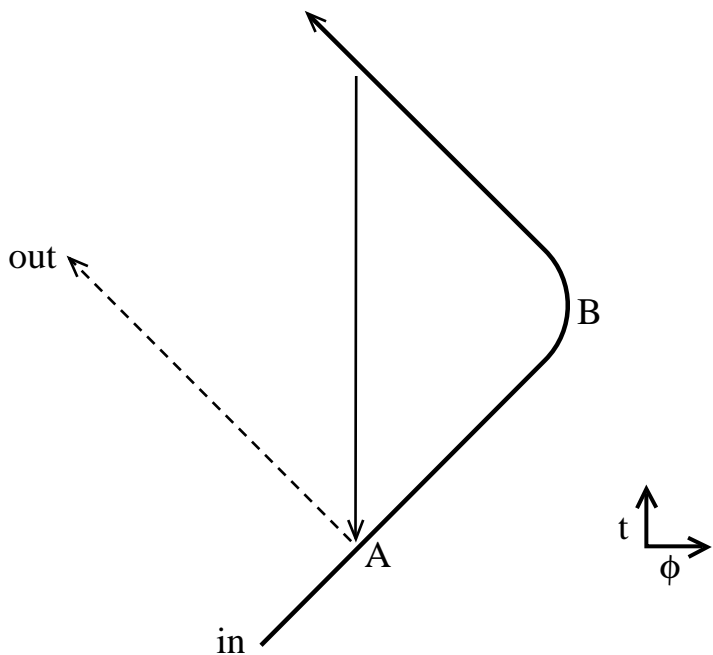}
\end{center}
\caption[]{Two tachyons in an incoming pulse can backscatter into
one (dashed line) at a given point~A.  In the matrix model all
tachyons pass through the turning region B.  The nonlocal
renormalization~(\ref{renorm}), vertical arrow, produces the
outgoing wave.}
\end{figure}
To analyze this it is
useful to introduce the conserved charges\cite{winf}
\begin{equation}
v_{mn} = e^{(n-m)t} \int_{F - F_0} \frac{dp\, dx}{2\pi}\, (-x-p)^m
(-x+p)^n , \label{vmn}
\end{equation}
with Poisson bracket algebra
\begin{equation}
\{ v_{mn}, v_{m'n'} \} = 2 (m'n - n'm) v_{m+m'-1,n+n'-1}.
\end{equation}
The phase space integral~(\ref{vmn}) runs over the interior $F$ of the
Fermi sea, minus the interior $F_0$ of the static sea,
$x < -\sqrt{p^2 + 2\overline\mu}$.
These are conserved because
\begin{equation}
(-x-p)e^{-t}, \qquad (-x+p)e^t \label{com}
\end{equation}
are constants of the motion in the inverted harmonic potential.
For perturbations which are not too large,
the Fermi sea can be described in terms of its upper and lower surfaces
$p_{\pm}(x,t)$.  Asymptotically, these move with velocity~1 in the
coordinate $q = - \ln (-x)$,
\begin{equation}
p_{\pm}(x,t)\ \sim\ \mp x \pm \frac{1}{x} \epsilon_{\pm}(t \mp q),
\label{lin}
\end{equation}
where the upper and
lower signs refer respectively to the incoming and outgoing waves.
The incoming and outgoing parts of the canonically normalized matrix
model scalar are related to the perturbation of the Fermi sea by
\begin{equation}
\partial_q \overline{\cal S}_{\pm}(t\mp q) = -\frac{1}{2\sqrt{\pi}}
\Bigl\{ \epsilon_{\pm}(t \mp q) - \overline\mu \Bigr\}. \label{cals}
\end{equation}
Evaluating the charge~(\ref{vmn}) in the limits $t \to \mp \infty$
gives a relation between the incoming and outgoing waves,
\begin{eqnarray}
v_{mn} &=& \frac{2^n}{2\pi(m+1)} \int_{-\infty}^\infty dt\,
e^{(n-m)(t-q)}
\Bigl\{ (\epsilon_+(t-q))^{m+1} - \overline\mu^{m+1} \Bigr\}
\nonumber\\
&=& \frac{2^m}{2\pi(n+1)} \int_{-\infty}^\infty dt\,
e^{(n-m)(t+q)}
\Bigl\{ (\epsilon_-(t+q))^{n+1} - \overline\mu^{n+1} \Bigr\}.
\label{vasym}
\end{eqnarray}

The $2 \to 1$ bulk scattering is now derived as follows.
To resolve the bulk scattering we use narrow wave packets as in
ref.~\cite{NP}, so the leading behavior as $t + \phi \to -\infty$
comes from the first pole in the renormalization~(\ref{renorm}),
at $\partial_\phi = 1$.  Near this point the renormalization factor
is $(\pi/2)^{-1/4} (\partial_\phi - 1)^{-1}$, giving
\begin{eqnarray}
\lim_{t + \phi \to -\infty} {\cal S}_{-}(t + \phi)
&=& -2^{1/4} \pi^{-1/4} e^t \int_{t}^\infty
dt'\, e^{- t'} \overline{\cal S}_{-}(t' + \phi) \nonumber\\
&=& -2^{1/4} \pi^{-1/4} e^t \int_{-\infty}^\infty
dt'\, e^{- t'} \overline{\cal S}_{-}(t' + \phi) \nonumber\\
&=& 2^{-3/4} \pi^{1/4} e^{t + \phi} v_{10}.
\end{eqnarray}
In the second line we have used the narrowness of the wavepacket to
extend the range of integration, and in the third we have noted that
the result is simply proportional to the conserved charge $v_{10}$.
Now expressing this in terms of the incoming field gives
\begin{eqnarray}
\lim_{t + \phi \to -\infty} {\cal S}_{-}(t + \phi)
&=& 2^{-3/4} \pi^{1/4} e^{t + \phi}
\int_{\infty}^\infty dt'\, e^{-t'+\phi}
\Bigl\{ (\partial_t \overline{\cal
S}_+(t-\phi))^2 + \overline\mu \partial_t\overline{\cal
S}_+(t-\phi) / \sqrt{\pi} \Bigr\} \nonumber\\
&=& 2^{-1/2} e^{t + \phi} \int_{\infty}^\infty dt'\,  e^{-t'+\phi}
({\cal S}_+(t'-\phi))^2 \ +\ O(\overline\mu)\ .
\label{2to1}
\end{eqnarray}
In the second line we have carried out the
renormalization~(\ref{renorm}) in reverse, leading to a simple result
after integration by parts.  The $O(\overline\mu)$ term
is from
$1 \to 1$ scattering on the tachyon background.

The result~(\ref{2to1}) is the same as in ref.~\cite{NP},
but the
derivation is actually more general.  Ref.~\cite{NP} used a
weak-field expansion, in powers of ${\cal S}$.  From the
definitions~(\ref{lin}) and~(\ref{cals}) the
weak-field expansion parameter is $\partial_\phi {\cal S}_+ / x^2$.
There is a second expansion parameter,
$1/\overline\mu^2$, governing the string loop expansion; for convenience we
assume this to be small and focus on classical scattering.
At large $x$, far in the asymptotic region,
$\partial_\phi {\cal S}_+ / x^2$ is always small,
but the condition that a pulse remain small throughout the
scattering is $\partial_\phi {\cal S}_+ \ll \overline\mu$ since
the turning radius is $O(\overline\mu^{1/2})$.  Ref.~\cite{NP}
required the latter inequality to hold; the derivation above does
not, because a conservation law has been used to relate the
incoming and outgoing waves.

This point is important, and a bit confusing, so let us expand
on it.  The bulk scattering process occurs in asymptotic region
where nonlinearities are always small, and so is determined entirely
by the cubic term in the effective Lagrangian, known from string
perturbation theory to be of the form $e^{2\phi} {\cal S}^3$.
This must be true even for a pulse with $\partial_\phi {\cal S}_+
\sim \overline\mu$.  For such a pulse the nonlinearities will
eventually become large, but this in a region (the turning region of
figure~1) in the future of the point where the bulk scattering
occurs, and so must be irrelevant.  However, the matrix model
completely obscures this causal relationship---the `information'
in the incoming pulses always propagates through the turning region
before being transferred to the outgoing wave by the nonlocal
renormalization.  The nonlinearity, even if strong, conserves
$v_{mn}$, so the correct amplitude is obtained just the same.

That is, all is well provided the pulse remains below the maximum of
the potential, which is to say in the region $x < -|p|$ or
$\partial_q \overline{\cal S}_{+} < \overline\mu /
2\sqrt{\pi}$. A pulse which exceeds this, shown in figure~2a,
will propagate to positive $x$ and, in the type~I theories,
feel the modification of the potential.
\begin{figure}
\epsfbox{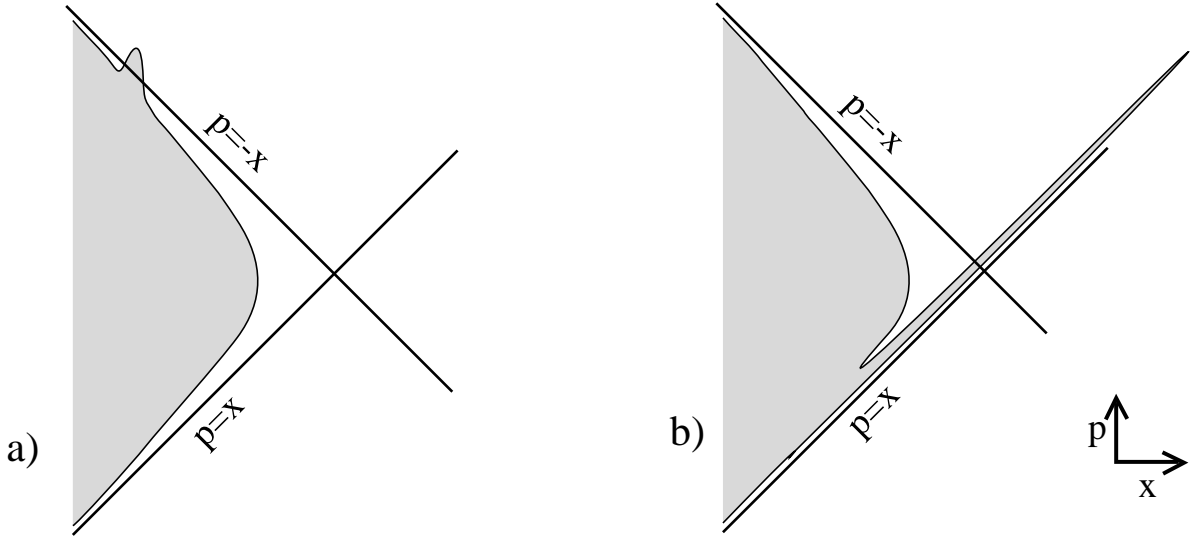}
\caption[]{a) Incoming pulse in phase-space plane, extending
above the line $p = -x$. Filled states are shaded.
b) Later form of the pulse, with part
propagating to $x = \infty$.}
\end{figure}
Consider for example
an infinite wall at $A = 0$: a fermion reaching the phase space
point $(x,p) = (0, p)$ will jump discontinuously to $(0, -p)$.
The quantities~(\ref{com}) change discontinuously and so the
$v_{mn}$ are not conserved.  The bulk scattering amplitude is
then {\it not} in agreement with string perturbation theory.
This is unacceptable, because the bulk scattering occurs in a
region where both the string and the weak-field expansion
parameters are small, while the strong nonlinearities occur
only in the causal future.  Moreover this is true for any
theory of type~I: the quantity $(-x-p)e^t$ is negative for
incoming fermions above the $p = - x$ line, but is positive for
all outgoing fermions.

The scattering process used in ref.~\cite{NP} to measure the
gravitational mass similarly depends upon conservation of
$v_{20}$.  As it happens, there is one theory of type~I which
conserves
$v_{20}$ and so the gravitational mass, but not $v_{10}$.
This is just the theory with infinite barrier at $x = 0$,
because the sign jump in $(0, p) \to (0, -p)$ cancels when
$m+n$ is even.  However, the causality violation in $2 \to 1$
scattering is still unacceptable.

Other bulk processes require conservation of $v_{m0}$ for all positive integer
$m$.  It is not clear whether a similar causality argument can be applied
to the $v_{mn}$ for $m$ and $n$ nonzero, but there is another way to see
that these must be conserved, and therefore that the type~I theories
are not consistent as string theories.
The $v_{mn}$ for $m, n \neq 0$ are just the unbroken symmetries of string
field theory discussed in ref.~\cite{WWZ},
with $v_{s+n,s-n}\ \propto\
A_{s,n}$, and so
should be conserved even nonperturbatively.\footnote
{The transformation of
the tachyon under the charges with $m=0$ or $n=0$
contains a $c$-number piece.  These
generators are thus not symmetries of the vacuum, and
do not appear as conserved currents in the
conformal field theory.

Only integer values of $m$ and $n$ appear in the
bulk scattering and as string field symmetries.  However, for a Fermi
sea entirely within the region $x < -|p|$, the charges $v_{mn}$ are well
defined for non-integer $m,n$ (provided the integrals converge).
The relation~(\ref{vasym}) between incoming and outgoing waves is then
the one found in ref.~\cite{MP} by other means.}

Even though we are discussing classical scattering this
condition is nonperturbative in the $1/\overline\mu$ expansion,
since the
condition for a large pulse is
$\overline\mu < 2\sqrt{\pi} \partial_q \overline{\cal S}_{+}.$
That is, the number of incoming tachyons is of order
$(1/\overline\mu)^{-2}$.

The other natural nonperturbative
definition of the theory, type~II, leaves the potential
unmodified.  Filling the Fermi sea to the same level
$-\overline\mu$ on both sides gives a stable state, and the matrix model
is a unitary theory with two asymptotic weak coupling
regions.
But even though the $v_{mn}$ are conserved
this is not a consistent string theory,
at least with the straightforward interpretation that each asymptotic
region of the matrix model corresponds to an asymptotic region of
spacetime.  The point is that a
part of the incoming fermion pulse travels over the barrier,
so that the relation~(\ref{vasym}) between $v_{mn}$ and the
left outgoing wave $\epsilon_-$ no longer holds and the correct
bulk scattering and gravitational mass are no longer obtained.

To summarize, both the type~II theory as defined above and the
infinite class of type~I theories have been previously assumed to
define consistent nonperturbative string theories.  We have argued
that in fact none of these are consistent, and it stands as a
challenge to find any consistent nonperturbative definition of the
$d=2$ string theory.  If one is to make sense of the type~II case,
it is evidently necessary to identify the asymptotic region of the
string theory with some combination of the two asymptotic regions
of the matrix model.  If one is to make sense of the type~I case,
the mapping~(\ref{renorm}) between the matrix model and string
Hilbert spaces must be modified.  This modification
will have to be rather complicated---the simple linear
relation~(\ref{renorm}) appears to be exact as long as the pulses
stay below the threshold, while past this point the mapping must
change abruptly.

The nonperturbative formulation of string theory is likely to
involve unexpected new ideas.  The matrix model is one of the few
clues available, and so it is important to resolve the issue raised here
and construct the exact theory.  We believe the correct formulation will
not involve modification of the potential, but will require a more subtle
mapping between the matrix model and string Hilbert spaces; this is under
active investigation.

One of the obstacles to progress in matrix models has been
that the number of consistent models---the number of
modifications and generalizations that one might try---is much
larger than the number of consistent string theories.  The
causality condition we have introduced is therefore a useful tool
in applying matrix models to critical string theory.  For example,
we have used it in ref.~\cite{NP} to argue that the proposed
matrix model black hole of ref.~\cite{JevYon} gives the wrong
bulk scattering at long distance (this point is also made in
refs.~\cite{BerKut} and~\cite{DemRod}).  As an aside, it seems very
likely that pulses which pass over the barrier as in figure~2 are
related to black hole formation, so the resolution of the problem
we have presented is also likely to lead to progress in this area.

The matrix model
might have been interpreted to give evidence for the existence of
a large number of nonperturbative parameters in string theory.
This would be unsatisfactory for the ultimate predictive power of
the theory, and rather surprising as well since all parameters in
string theory are believed to be associated with background fields. Our
result is thus further evidence for the uniqueness of string theory.

\section*{Acknowledgements}

I would like to thank Shyamoli Chauduri and Makoto Natsuume for
discussions.  This work was supported in
part by National Science Foundation grants PHY-89-04035 and
PHY-91-16964.

\end{document}